\documentclass[11pt,preprint,apjfonts]{aastex}
\usepackage{emulateapj5,psfig}

\def\gsim{\mathrel{\raise0.35ex\hbox{$\scriptstyle >$}\kern-0.6em
\lower0.40ex\hbox{{$\scriptstyle \sim$}}}}
\def\lsim{\mathrel{\raise0.35ex\hbox{$\scriptstyle <$}\kern-0.6em
\lower0.40ex\hbox{{$\scriptstyle \sim$}}}}

\newenvironment{inlinefigure}{%
\def\@captype{figure}%
\noindent\begin{minipage}{0.999\linewidth}\begin{center}}
{\end{center}\end{minipage}\smallskip}
\makeatother

\begin{document}

\title{A Detailed Study of Gas and Star Formation in a Highly 
Magnified Lyman Break Galaxy at $z=3.07$}

\lefthead{CO and MIR Emission in the Cosmic Eye}
\righthead{Coppin et al.}

\author{K.\,E.\,K.\ Coppin\altaffilmark{1}, 
A.\,M.\ Swinbank\altaffilmark{1}, 
R.\ Neri\altaffilmark{2}, 
P.\ Cox\altaffilmark{2},
Ian Smail\altaffilmark{1},
R.\,S.\ Ellis\altaffilmark{3},\\
J.\,E. Geach\altaffilmark{1},
B.\ Siana\altaffilmark{4}, 
H.\ Teplitz\altaffilmark{4}, 
S.\ Dye\altaffilmark{5},
J.-P.\ Kneib\altaffilmark{6}, 
A.\,C. Edge\altaffilmark{1},
J.\ Richard\altaffilmark{3}\\
}

\setcounter{footnote}{0}

\altaffiltext{1}{Institute for Computational Cosmology, Durham University, 
South Road, Durham, DH1 3LE, UK -- Email: kristen.coppin@dur.ac.uk}
\altaffiltext{2}{Institut de RadioAstronomie Millim\'{e}trique (IRAM), 
300 rue de la Piscine, Domaine Universitaire, 38406 Saint 
Martin d'H\`{e}res, France}
\altaffiltext{3}{Caltech, MC 105-24, 1200 East California Blvd, Pasadena, 
California, CA91125, USA }
\altaffiltext{4}{Spitzer Science Center, Caltech, MC 314-6, 1200 East 
California Blvd, 
Pasadena, California, CA91125, USA}
\altaffiltext{5}{School of Physics and Astronomy, Cardiff University,
  5, The Parade, Cardiff, Wales, CF24AA, UK}
\altaffiltext{6}{Laboratoire d'Astrophysique de Marseille, Traverse 
du Siphon -- B.P.8 13376, Marseille Cedec 12, France}
\setcounter{footnote}{6}
\setcounter{figure}{0}

\begin{abstract}
We report the detection of CO(3--2) emission from a bright,
gravitationally lensed Lyman Break Galaxy, LBG\,J213512.73--010143 (the
``Cosmic Eye''), at $z=3.07$ using the Plateau de Bure Interferometer.
This is only the second detection of molecular gas emission from an
LBG and yields an intrinsic molecular gas mass of $(2.4
\pm 0.4)\times 10^{9}$\,M$_\odot$.
The lens reconstruction of the UV morphology of the LBG indicates that
it comprises two components separated by $\sim 2$\,kpc.  
The CO emission is unresolved, $\theta \lsim 3\arcsec$, and 
appears to be centered on the intrinsically fainter 
(and also less highly magnified)
of the two UV components.  The width of the CO line indicates a dynamical
mass of $(8\pm 2) \times 10^{9}{\rm csc}^2 i$~M$_\odot$ within the central
2\,kpc.  Employing mid-infrared observations from {\it Spitzer} we 
infer a stellar mass of
$M_{*}\sim (6\pm2)\times$10$^9$\,M$_\odot$ and a 
star-formation rate of $\sim60$\,M$_\odot$\,yr$^{-1}$, indicating
that the molecular gas will be consumed in $\lsim40$\,Myr.  The gas
fractions, star-formation efficiencies and line widths suggests that
LBG\,J213512 is a high-redshift, gas-rich analog of a local luminous
infrared galaxy.  This galaxy has a similar gas-to-dynamical mass
fraction as observed in the submillimeter-selected population, although
the gas surface density and star-formation efficiency is a factor of
$3\times$ less, suggesting less vigorous activity.  We 
discuss the uncertainties in our conclusions arising from adopting a
CO-to-H$_2$ conversion factor appropriate for either the Milky Way or local
luminous infrared galaxies.  These observations demonstrate that
current facilities, when aided by fortuitous gravitational
magnification, can study ``ordinary'' galaxies at high-redshift and so
act as pathfinders for ALMA.
\end{abstract}


\section{Introduction}

Lyman Break Galaxies (LBGs) were the first significant population of
high-redshift galaxies to be identified \citep{Steidel96}.  
Subsequent work has shown
that they represent only a subset of the galaxy population at $z\sim3$,
whose properties in part reflect their selection: LBGs are actively
star-forming galaxies with relatively low dust obscuration.  However,
LBGs are still the most common population at this epoch and as such
they have been interpreted as a phase in the formation of ``typical''
galaxies (\citealt{Somerville01}; \citealt{Baugh05}).  Thus, understanding 
their properties, such as the distribution of star-formation, dynamical, 
stellar and gas masses, may be a critical element in constraining 
models for the formation and evolution of normal galaxies.

Studies of the physical properties of LBGs have so far traced their
star-formation rates and histories, stellar and 
dynamical masses and morphologies 
(e.g.~\citealt{Shapley01,Shapley03,Shapley05}; 
\citealt{Reddy04}; \citealt{Law07}).  These studies have motivated more
detailed investigations of the properties of LBGs, in particular to
determine the gas content of these galaxies and the chemical enrichment
of this gas, to more fully understand their evolutionary status.  In
particular, the gas content of these galaxies (as traced by their CO
emission in the millimeter waveband) is a key observable, as this 
cold and dense gas
provides the reservoir for star-formation activity (and hence the
potential to build up a substantial stellar mass).  CO emission 
provides both a reliable measure of the gas mass and also 
an unbiased tracer of its dynamics and hence the mass of
the host galaxy.

However, the properties of the gaseous component in typical LBGs have
proved hard to address, as they are beyond the sensitivity limits of
the relevant current observational facilities.  Progress has so far only been
made through studies of a single, rare example of an LBG whose apparent
brightness is boosted by gravitational magnification by a foreground
cluster lens: MS 1512--cB58 (hereafter referred to as cB58:
\citealt{Yee96}).  cB58 has been the subject of two unique studies of
the interstellar medium of LBGs: \citet{Pettini00,Pettini02} derived the
elemental abundances from high signal-to-noise (S/N), high-resolution 
restframe UV spectroscopy
concluding that cB58 is a $Z\sim 0.5Z_{\odot}$ starburst galaxy. While
\citet{Baker04} obtained for cB58 the only detection of CO emission
from an LBG, providing the first direct evidence of the existence of a
sizeable cold gas reservoir in an LBG.  These observations have also 
helped to shed light
on several key details of the star-formation process in this young galaxy.

Unfortunately, while these studies of cB58 have provided unique
insights into the properties of LBGs, it is dangerous to draw general
conclusions about the whole LBG population from this single example.
Hence, significant efforts have gone into finding other examples of
lensed LBGs and after nearly a decade of searches, several lensed LBGs
as bright as cB58 have now been found.  \citet{Smail06} present the
discovery of a strongly lensed LBG at $z=3.07$, LBG J213512.73--010143
(hereafter LBG\,J213512), similar to cB58.  At $r_\mathrm{AB}=20.3$
this new LBG is brighter in the restframe UV than cB58, owing to the
$28\times$ magnification \citep{Dye}.  Moreover, the LBG appears as two
small arcs ($\sim3\arcsec$ in extent) and hence provides a unique
opportunity for spatially resolved studies on 100\,pc scales across
this galaxy.  When corrected for the lensing magnification, the
background source appears as an $L^\star$ LBG, with a similar intrinsic
luminosity to cB58.  We thus have a second example of a highly
magnified ``typical'' LBG.

In this paper we present interferometric measurements of the CO(3--2)
emission and mid-infrared photometric observations 
of this new lensed LBG.  We use
the CO(3--2) line luminosity and width to infer the gas and dynamical
masses and compare these with the stellar mass inferred from the
rest-frame optical-to-near-infrared photometry of the system.  We also
compare our results with the similar observations of cB58 to
investigate the variations within the LBG population and further
compare our results with the growing number of CO observations of
other high-redshift galaxies (see \citealt{Solomon_rev} for a review).
We adopt cosmological parameters from the \textit{WMAP} fits 
in \citet{Spergel}: $\Omega_\Lambda=0.73$, $\Omega_m=0.27$, and
$H_\mathrm{0}=71$\,km\,s$^{-1}$\,Mpc$^{-1}$.  All quoted magnitudes are
on the AB system.

\section{Observations and Reduction}

\subsection{Millimeter Interferometry}

We used the six-element IRAM PdBI \citep{Guilloteau} to observe
LBG\,J213512 in the redshifted CO(3--2) line and in the continuum at
84.87\,GHz.  The frequency was tuned to the CO(3--2) rotational
transition at $z=3.0743$, the systemic redshift of the system from
\citet{Smail06}.  Observations were made in D configuration in
Director's Discretionary Time (DDT) between 2006 August 24 and 2006
September 20 with good atmospheric phase stability (seeing =
0.6$\arcsec$--1.6$\arcsec$) and reasonable transparency (pwv =
5--15\,mm).  We observed LBG\,J213512 with a total on-source observing
time of 10\,hrs.  The spectral correlator was adjusted to detect the
line with a frequency resolution of 2.5\,MHz in the 580\,MHz band of
the receivers. The overall flux scale for each observing epoch was set
on 2134+004, 2145+067 and MWC349.  The visibilities were resampled to a
velocity resolution of 35.3\,km\,s$^{-1}$ (10\,MHz) providing 1-$\sigma$ line
sensitivities of $\sim1.0\,\mathrm{mJy\,beam^{-1}}$. 
The corresponding
synthesized beam, adopting natural weighting, was
6.3$\arcsec\times$\,5.5$\arcsec$ at $44^\circ$ east of north. The data
were calibrated, mapped and analyzed in the {\sc gildas} software
package.  Inspection of the velocity datacube shows a significant
detection of CO(3--2) line emission close to the position of
LBG\,J213512 in the central velocity channels with a velocity width of
$\sim250$\,km\,s$^{-1}$.  Fig.~1 shows a channel map constructed from
the average emission in a 250\,km\,s$^{-1}$ window centered on the
systemic redshift of the system.  In Fig.~2 we show the 
spectrum of the CO(3--2) emission in the brightest pixel of the source
in the channel map.

\begin{inlinefigure}
  \centerline{\psfig{file=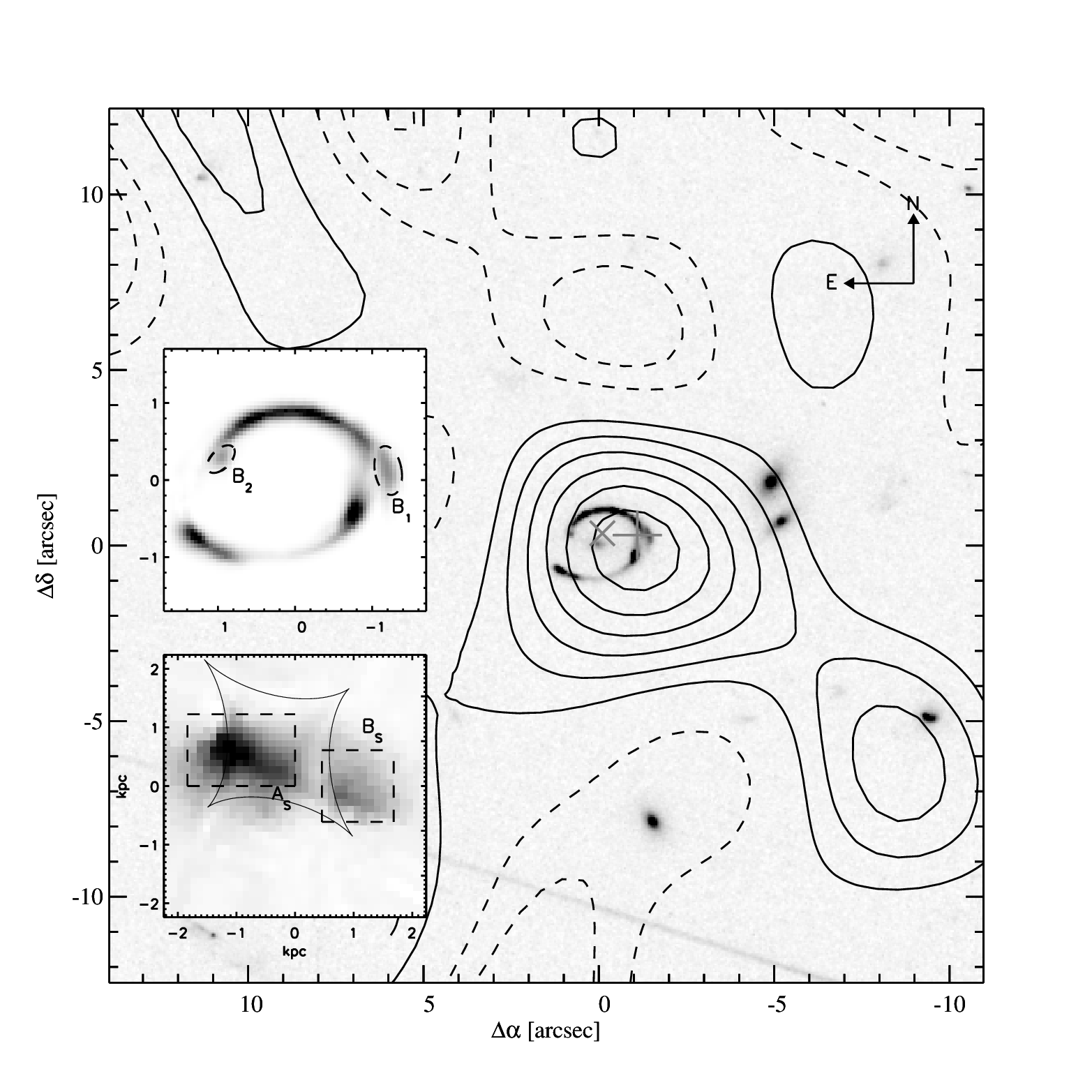,width=3.5in,angle=0}}
  \figcaption{{\it HST} ACS F606W image of LBG\,J213512, centered on
    the optical position, 21\,35\,12.73 $-01$\,01\,43.0 (J2000), with
    the CO(3--2) contours overlaid (contours begin at $1\sigma$ and
    increase in steps of $1\sigma$, negative contours are dashed).  We
    also show two insets, the upper inset illustrates the image-plane
    morphology of the LBG (with the foreground lens removed; scale in
    arcseconds) consisting of two bright arcs and two fainter knots
    (marked B$_1$ and B$_2$).  The lower inset illustrates the
    source-plane reconstruction of the LBG from \citet{Dye} in kpc.
    This shows that in the restframe UV the LBG comprises two
    components (A$_\mathrm{s}$ and B$_\mathrm{s}$) separated by
    $\sim$\,2--3\,kpc.  Both of the bright arcs are formed from the
    lensing of A$_\mathrm{s}$, which is magnified in total by a factor
    of $28\times$, while B$_\mathrm{s}$ gives rise to B$_1$ and B$_2$
    (the magnifications of $B_{1}$ and $B_{2}$ are 6.2$\times$ and
    1.8$\times$, respectively).  Comparison of the CO map with the
    optical morphology of the system shows the CO emission is offset
    from the center of the LBG, and we mark by a ``$\times$'' the
    predicted center of the CO emission if the molecular gas follows
    the $R$- or $K$-band light.  We conclude that the CO emission is
    unlikely to follow the restframe UV/optical light in this system
    and instead it appears to coincide with a faint knot (marked as
    B$_1$ in the top inset).  To test this we predict the observed
    centroid of the CO emission if it is associated with source
    B$_\mathrm{s}$ and mark this as ``+'', this is coincident with the
    observed centroid supporting the proposed association.  See text
    for a detailed discussion.}
\label{contour}
\end{inlinefigure}
\addtocounter{figure}{0}

\subsection{Mid-infrared Imaging}

Observations of LBG\,J213512 were taken with the \textit{Spitzer Space
Telescope} Infrared Array Camera (IRAC) in 2005 November 
using time awarded through DDT.
Observations were taken at 3.6, 4.5, 5.8 and 8.0\,$\mu$m using two
cycles of a 10-point dither pattern, each with integration time of
30\,s.  The total integration time was 0.9\,ks\,pixel$^{-1}$ in each
band.  Further observations were also obtained at 24\,$\mu$m with the
Multi-band Imaging Photometer for \textit{Spitzer} (MIPS) camera.
These observations were made using eight repeats of a nine-point dither
pattern, each consisting of 30\,s exposures, for a total integration
time of 2.0\,ks.  We use the Post Basic Calibrated IRAC frames
generated by the pipelines at the \textit{Spitzer} Science Center (SSC)
and perform post-processing on the data frames to remove common
artifacts and flatten small and large-scale gradients using
``master-flats'' generated from the data. For mosaicking we use the SSC
{\sc mopex} package, which makes use of the supplementary calibration
files which are supplied with the main science set.  We align and
combine the four IRAC images and use this composite image to extract
$3.8\arcsec$-diameter apertures magnitudes at the position of LBG\,J213512
from the IRAC data and $10\arcsec$-diameter aperture magnitudes 
from the MIPS data. We obtain a strong detection of the emission from the 
LBG\,J213512 (with some contribution from the foreground lens) at
3.6--8$\mu$m and a weaker detection at 24\,$\mu$m.
Aperture-corrected photometry for the LBG are reported in Table~1, 
including $gR_{606}K$-band photometry from \citet{Smail06}.

\begin{center}
\small
\centerline{\sc Table 1.}
\centerline{\sc Aperture-corrected photometry for  LBG\,J213512}
\medskip

\begin{tabular}{lccc}
\hline\hline
\noalign{\smallskip}
           & LBG + Lens      & Lens      & LBG   \\  
           & (1)            & (2)       & (3)  \\
\hline
\noalign{\smallskip}
$g$          & \nodata  & \nodata   & 21.47[5]    \\
$R_{606}$  & \nodata  & 22.34[15] & 20.54[2]     \\
$K$          & \nodata & 19.70[10] & 18.90[10]      \\
3.6\,$\mu$m  & 18.34[2]  & \nodata   & 18.74[4]       \\
4.5\,$\mu$m  & 18.19[2]  & \nodata   & 18.39[4]    \\
5.8\,$\mu$m  & 18.13[3]  & \nodata   & 18.33[6]     \\
8.0\,$\mu$m  & 18.30[2]  & \nodata   & 18.35[4]     \\
24\,$\mu$m   & 17.78[25] & \nodata   & 17.78[25]  \\
\noalign{\smallskip}
\hline\hline
\label{tab:photom}
\end{tabular}
\end{center}
\medskip
\begin{minipage}{3.5in}
  \small {\sc Note.} -- (1) The observed photometry for the LBG.  Due to the low spatial resolution of IRAC the photometry from the $z$=0.73 and $z$=3.07 galaxies are blended.  We remove the contribution of the $z$=0.73 lensing galaxy to the IRAC channels by fitting the $R_{606}\&K$-band photometry of the foreground lens (column 2) with an elliptical galaxy template redshifted to $z$=0.73 and subtracting the contribution at $\geq3.6\,\mathrm{\mu m}$.  These corrected LBG photometry are given in column 3 (see text).  To convert to intrinsic magnitudes (corrected for the lensing amplification factor $28\pm{3}\times$) add 3.6\,mags to column 3.  Values in the [] denote the error in the last decimal place and we note that 1\,$\mu$Jy corresponds to m$_{\rm AB}=23.90$.
\end{minipage}

\section{Analysis and Results}

\subsection{CO Emission Properties}

The luminosity, velocity width and spatial extent of the CO line
emission can be used to place limits on the gas and dynamical mass of
the system.  Fitting a Gaussian profile to the CO spectrum for the
source at the phase center of our observations (Fig.~2) we derive a
best-fit redshift for the CO(3--2) emission of $z=3.0740 \pm 0.0002$.
This is in excellent agreement with the systemic redshift derived from
the [O{\sc iii}]\,$\lambda$5007 line by \citet{Smail06}, with an offset
of just $\Delta\,v=-22\pm 14$\,km\,s$^{-1}$.  The error on the line
parameters are bootstrap estimates generated by adding noise at random
to the channels closest to the spectral peak from the outer-most noisy
baseline channels $1000\times$ with replacement (e.g.\ see
\citealt{Wall}).  This provides a reasonable estimate of the
uncertainties provided the noise is not strongly correlated between
channels. The restframe FWHM of the CO line is $190\pm
24$\,km\,s$^{-1}$ (or alternatively a dispersion of
$80\pm10$\,km\,s$^{-1}$) which is consistent with the [O{\sc
  iii}]\,$\lambda$5007 emission line
($\mathrm{FWHM}\lsim 220\,\mathrm{km\,s^{-1}}$) as measured by
\citet{Smail06}.  The velocity integrated line flux is measured to be
$F_\mathrm{CO(3-2)}=0.50\pm{0.07}\,\mathrm{Jy\,km\,s^{-1}}$ (S/N=7.1).
We note that this line flux is comparable to the CO flux density
measured for cB58:
$F_\mathrm{CO(3-2)}=0.37\pm{0.08}\,\mathrm{Jy\,km\,s^{-1}}$
(\citealt{Baker04}, with an estimated magnification factor for cB58 of
$\mu\sim32$).

\begin{inlinefigure}
  \centerline{\psfig{file=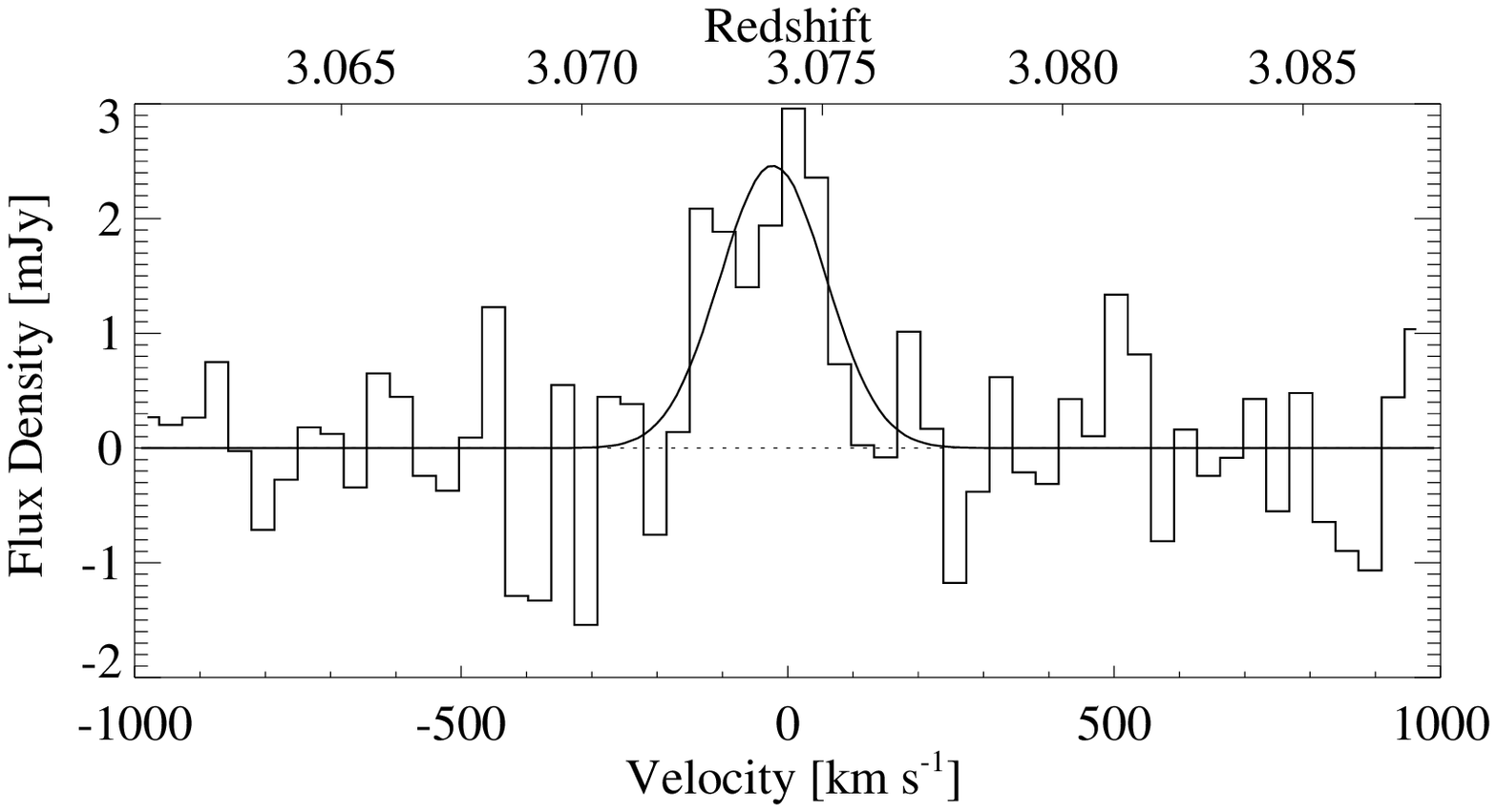,width=3.0in,angle=0}}
  \figcaption{The spectrum of CO(3--2)
    emission in LBG\,J213512, binned into 35.3\,km\,s$^{-1}$ channels.
    We detect the line at 7.1-$\sigma$ significance and the 
best-fitting Gaussian with a FWHM of $190\pm24$\,km\,s$^{-1}$
    is overlaid.  The upper axis indicates the redshift
    of the CO(3--2) transition. There is a hint of double peak to the
    CO line, although a double component fit does not provide a better
    description of the data due to the modest S/N of the detection.  
We have confirmed that there is no detectable spatial
    offset between the blue and red halves of the line ($\lsim
    0.1\arcsec$).}
\label{fig:spectrum}
\end{inlinefigure}
\addtocounter{figure}{0}

Next, averaging the emission over a 250\,km\,s$^{-1}$ window, centered
on the redshift of the CO emission, we determine a flux of
$2.0\pm0.3$\,mJy\,beam$^{-1}$ for a source centered at 21\,35\,12.62, 
$-01$\,01\,43.9 (J2000) corresponding to a
$\sim 7$-$\sigma$ detection (we ignore the small correction for
primary beam attenuation given the source is close to the phase center
of the map).  We note that no significant continuum emission is
detected from the line-free region (470\,MHz of bandwidth) down to a
1-$\sigma$ sensitivity of 0.14\,mJy.

In Fig.~1 we show the {\it Hubble Space Telescope (HST)} ACS F606W
image of the system from \citet{Smail06} and overlay the contours from
the CO(3--2) channel map.  \citet{Dye} present a detailed lens model
and use it to reconstruct the restframe UV source-plane morphology of the LBG.
They show that the system comprises two UV-emitting components separated by
$\sim2$--3\,kpc, probably representing two star-forming regions 
within the galaxy separated 
by a dust lane or region of lower activity (see inset in Fig.~1; 
A$_\mathrm{s}$ and
B$_\mathrm{s}$).  The intrinsically brighter and more highly amplified
of these, A$_\mathrm{s}$, lies within the lens caustic and is magnified
by a factor of $28\pm3$ \citep{Smail06}, giving rise to the ring-like
structure surrounding the foreground galaxy.  The second UV component,
B$_\mathrm{s}$, lies just outside the caustic and gives rise to two
images: B$_{1}$ and B$_{2}$ (see Fig.~1 and
Fig.~4 of \citealt{Dye}).  The lensing magnifications of the components
are $6.2\times$ and $1.8\times$ respectively, giving a total
amplification factor of $(8.0 \pm 0.9)\,\times$ for B$_\mathrm{s}$ \citep{Dye}.

The astrometric solution on the {\it HST} image is derived by comparing
the positions of unsaturated stars with the USNOA-2.0 catalog and we
estimate an r.m.s.\ uncertainty in the optical position of the system
(21\,35\,12.73, $-01$\,01\,43.0, J2000) of $\lsim 0.2''$.  The absolute
astrometry of the CO observations is estimated to be accurate to
$\lsim0.01''$, but the centroid for the emission is only accurate to
$\lsim 0.3''$ and places the peak of the CO emission at 21\,35\,12.62,
$-01$\,01\,43.9 (J2000) -- which is centered approximately $1.9\pm
0.4\arcsec$ due west of the optical centroid.

We first test whether the CO centroid is consistent with the expected
position if the CO emission traces the restframe UV/optical light.  To
achieve this we subtract the foreground lensing galaxy from the {\it
  HST} imaging and convolve the resulting image with the PdBI beam,
this predicts a centroid for the CO emission approximately
$0.4\arcsec$ NW of the lensing galaxy (marked by the ``$\times$'' in
Fig.~1).  Thus the CO line emission does not appear to be associated
with the most highly magnified and UV-bright component part of the
galaxy (A$_\mathrm{s}$).  Instead, the centroid of the CO emission
appears to correspond to B$_1$, the more magnified of the two images
of component B$_\mathrm{s}$.  To confirm this, we model the CO
emission from B$_\mathrm{s}$ using the magnifications for the two
images from \citet{Dye} and convolve this with the PdBI beam, this
yields a centroid of 21\,35\,12.63, $-01$\,01\,43.3 (J2000) (marked
by the `+' in Fig.~1), which is $0.6\pm 0.4\arcsec$ due west from the
observed peak and hence consistent with it.  Thus, we conclude that
the CO reservoir appears to be associated with B$_\mathrm{s}$, the
fainter of the two UV regions in LBG\,J213512.  We discuss the
implications of this further in \S4.

\subsection{CO Luminosity and Molecular Gas Mass}

We calculate the line luminosity and
estimate the total cold gas mass (H$_{2}$+He) from the integrated CO line flux.
The observed CO(3--2) line luminosity is
$\mu\,L'_\mathrm{CO(3-2)}=(2.4\pm0.3)\times10^{10}\,\mathrm{K\,km\,s^{-1}\,pc^{2}}$,
where $\mu$ is the amplification factor.  Although the lensing
magnification of the entire UV source is $\sim 28 \times$, our
association of the molecular gas reservoir with component
B$_\mathrm{s}$ means we must adopt the estimated magnification for this
component of $(8.0 \pm 0.9)\,\times$.  The resulting intrinsic line
luminosity is therefore $L'_\mathrm{CO(3-2)}=(3.0
\pm0.5)\times10^{9}\,\mathrm{K\,km\,s^{-1}\,pc^{2}}$.  We then assume
both a line luminosity ratio of
$r_{32}=L'_\mathrm{CO(3-2)}/L'_\mathrm{CO(1-0)}=1$ 
(i.e. a constant brightness temperature) and a CO-to-H$_2$
conversion factor of
$\alpha=0.8\,$M$_{\odot}(\mathrm{K\,km\,s^{-1}\,pc^{2}})^{-1}$.  These
values are appropriate for local galaxy populations exhibiting similar
levels of star formation activity to LBG\,J213512 (e.g.\ 
local infrared galaxies or LIRGs; 
\citealt{S097}; see \S3.4).
We discuss in \S4.2 how our conclusions would
change if we adopted $\alpha$ and $r_{32}$ values typical of the Milky
Way.  This then yields a total cold gas mass of
M$_\mathrm{gas}=$\,M$($H$_2+$He$)=\alpha\,L'_\mathrm{CO}=(2.4\,\pm
0.4)\times 10^{9}$\,M$_{\odot}$.  Based on the association of the CO
emission with source B$_\mathrm{s}$ and its extent in the restframe UV
we assume the gas is distributed in a disk with a radius 
no larger than 1\,kpc (see
Fig.~1) resulting in an inferred gas surface density of
$\Sigma_\mathrm{gas}\simeq (760\pm130)\,$M$_{\odot}\,\mathrm{pc}^{-2}$.

\subsection{Dynamical and Stellar Mass}

The next step in our analysis is to compare the mass we derived for
the cold gas reservoir to the dynamical and stellar masses in this
LBG.  Whilst in principle the dynamical masses of LBGs can be derived
from optical or near-infrared observations of emission line gas in
these galaxies, dust obscuration and outflows in the gas may bias the
measurements.  In contrast, molecular CO emission is comparatively immune
to the effects of obscuration and outflows and therefore provides a
unbiased measurement of dynamics within the CO emitting region.  Our
CO observations allow us to place strong constraints on the dynamical
mass, while the {\it Spitzer} observations can be used to constrain
the stellar mass and current star formation rate (SFR) of the galaxy.

For LBG\,J213512, the line width of the CO emission ($190 \pm
24\,\mathrm{km\,s^{-1}}$) predicts a dynamical mass of $(8.4 \pm1.4)
\times10^{9}\,{\rm csc}^{2}i$\,M$_\odot$, assuming the gas lies in a disk
with inclination $i$ and a radius of $\sim 1$\,kpc.   Based on this
we calculate a gas-to-dynamical mass fraction of
$f=M_\mathrm{gas}/M_\mathrm{dyn}\sim0.30 \sin^2i$.
For comparison with the total stellar mass, we extrapolate this
CO measurement slightly to give us a total mass within the UV-extent of
the LBG, a radius of 2\,kpc, yielding an enclosed mass of
M$(<2$\,kpc$)\sim1.7\times 10^{10}\,{\rm csc}^2i$\,M$_\odot$.  
We note that 
the mean angle of randomly oriented disks with respect to the
sky plane in three dimensions is $i=30$\,deg \citep{CarWangErr}, 
resulting in an inclination correction of ${\rm csc}^2 i = 4$.

In order to derive the stellar mass, we adopt the approach of 
\citet{Borys05}: first we estimate the rest-frame
$K$-band luminosity for the galaxy by interpolating the rest-frame
spectral energy distribution (SED) and then combine this with estimates
of the $K$-band light-to-mass ratio (L$_K$/M) for a range of plausible
ages for the dominant stellar population.  We exploit the {\small{\sc
    HyperZ}} package \citep{hyperz} to fit the observed $gR_{606}K$ and
IRAC 3.6--8.0\,$\mu$m photometry in Table~1.  We use solar metallicity
stellar population models from \citet{BruzChar93} and a
\citet{Calzetti00} starburst attenuation to infer the rest-frame
optical/near-infrared SED.  At $z=3.07$, the rest-frame $K$-band (which
is most sensitive to the underlying stellar population) is redshifted
to $\sim 8\,\mathrm{\mu m}$ (i.e.\ into IRAC Channel 4).  However,
since the IRAC photometry has a spatial resolution of $\sim 3\arcsec$
comparable to the extent of the lensed LBG, the contribution from the
$z=0.73$ lensing galaxy is blended with the LBG, and the lensing galaxy
must be subtracted.  We therefore fit the $R_{606}$ and $K$-band
magnitudes of the lens with an early-type (E/S0) galaxy template 
based on the spectral 
properties and colors in \citet{Smail06} (the
template comprises a stellar population for an exponential
star-formation history with a timescale of 1\,Gyr) redshifted to
$z=0.73$ and subtract the expected contribution to the IRAC photometry:
$\lsim 20$, 15, 10 \& 5 percent at 3.6, 4.5, 5.0 and 8.0\,$\mu$m,
respectively.  The LBG photometry with the contribution from the
lensing galaxy removed are given in Table~1.

We determine a rest-frame absolute $K$-band magnitude of
$\mathrm{M}_{K}=-22.2 \pm 0.1$ (corrected for lens magnification using
$\mu=28$).  As expected, this is comparable to L$^{*}_{8_\mathrm{\mu
    m}}$ for $z\sim3$ LBGs \citep{Shapley05}.  To convert this to a
stellar mass, we need to determine L$_{K}$/M for the dominant stellar
population.  To do this, we turn to the {\sc Starburst99} stellar
population model \citep{Leitherer99}, which provides estimates for
L$_{K}$/M for models of bursts of star-formation.  We adopt an age for
the stellar population dominating the UV light of $\sim 10$--30\,Myr 
based on the analysis of
the features in the restframe UV spectrum which show that the UV
emission is dominated by B stars (\citealt{Smail06}).  Fits
using {\sc hyperz} \citep{hyperz} to the broadband SED spanning
the restframe 0.13--2\,$\mu$m yield similarly young ages, $\lesssim
50$\,Myr with moderate reddening, $A_V\sim 1$, as do more sophisticated 
modeling \citep{Shapley05}.  For these ages
the {\sc Starburst99} models predict L$_{K}/\mathrm{M} \sim 2.5\pm
0.5$.  We note that the average L$_{K}/\mathrm{M}$ for LBGs in 
\citet{Shapley05} is 
L$_{K}/\mathrm{M}\sim2.5$.  Thus for a stellar population dominated 
by young stars 
($<50$\,Myr), the predicted stellar mass is $\sim (6\pm 2) \times
10^{9}$\,M$_{\odot}$.  This implies a baryonic mass of
M$_\mathrm{bary}=$\,M$_\mathrm{gas}+$\,M$_\mathrm{stars}\lsim
1\times10^{10}$\,M$_{\odot}$, with 75 per cent of this in stars.
This baryonic mass estimate is  consistent with the dynamical mass 
with a 1--2-kpc radius traced by the CO emission: 
M\,$\sim0.8$--$1.7\times 10^{10}\,{\rm csc}^2i$\,M$_\odot$, for all
inclinations.

\subsection{Star-Formation Rate and Efficiency}

\citet{Smail06} estimate the SFR for LBG\,J213512
from the rest-frame 1500\AA\ continuum flux of
L$_{1500}\sim4.6\times10^{30}$\,erg\,s$^{-1}$Hz$^{-1}$ which translates
into a SFR (corrected for lens magnification and dust reddening) of
$\sim100\,$\,M$_{\odot}\,\mathrm{yr}^{-1}$ (assuming a Salpeter initial
mass function (IMF) with an upper mass cut off of 100\,M$_{\odot}$;
\citealt{Kennicutt98}).

Perhaps more reliably, the 24-$\mu$m flux of the LBG can be used to
estimate the far-infrared luminosity of the galaxy
(e.g.~\citealt{Bell05}; \citealt{Geach06}) which can then be converted
into a SFR.  Taking the library of dusty, star-forming SEDs from
\citet{DaleHelou02}, we calculate the ratio of the 24-$\mu$m luminosity
to the far-infrared luminosity integrated over the entire SED
(8--$1000\mu$m). The SED templates are simply characterized by a
power-law distribution of dust mass, with an exponent of 1--2.5 for
``typical'' local star-forming galaxies \citep{Dale01}.  We derive
L$_\mathrm{FIR}\sim3.4\times10^{11}$\,L$_\odot$ for an
unlensed 24-$\mu$m flux of 10\,$\mu$Jy, corresponding to a SFR of
$\sim60$\,M$_\odot\mathrm{yr}^{-1}$ \citep{Kennicutt98}.  This is the
average luminosity over the entire library of SEDs, and the extreme
SEDs suggest our estimate is likely to be uncertain by a factor of
$3\times$.  This bolometric luminosity is consistent with the 0.14\,mJy upper
limit to the 3\,mm flux of LBG\,J213512 and the luminosity implies an 
$850\,\mathrm{\mu m}$ flux of $\sim0.35$\,mJy, where we have estimated 
this flux by scaling a modified blackbody with a dust temperature 
of 40\,K, and a dust emissivity 
$\beta=1.5$ (see \citealt{Blain02}) to match the 
$24\,\mathrm{\mu  m}$-predicted bolometric luminosity. 
Note that $S_{850}\sim 0.35$\,mJy
is consistent with a mean observed $850\,\mathrm{\mu m}$ flux
of $0.5\,\pm{0.4}\,\mathrm{mJy}$ for typical LBGs \citep{Chapman00}.  
Our SFR estimate yields a SFR density
of $\Sigma_\mathrm{SFR}\sim20$\,M$_\odot$\,yr$^{-1}$\,kpc$^{-2}$ for a
disk radius of 1\,kpc, and a star-formation efficiency (SFE) of L$_{\rm
  FIR}/$M$_\mathrm{H_{2}}=3.4\times 10^{11}L_\odot/2.4\times
10^{9}$\,M$_\odot\,\sim140$\,L$_\odot$\,M$_\odot^{-1}$.

A related way to present these observations is to roughly estimate 
the gas-to-dust mass ratio.  
Following \citet{Blain02}, we estimate a dust mass of 
M$_\mathrm{d}\simeq2.5\times10^{7}$\,M$_\odot$ from
the predicted $850\,\mathrm{\mu m}$ flux of $\sim0.35$\,mJy, 
assuming a dust mass absorption coefficient of 
$\kappa_{850_\mathrm{\mu m}}=0.15\,\mathrm{m}^{2}\,\mathrm{kg}^{-1}$.  
This yields a constraint on
the gas-to-dust mass ratio for LBG\,J213512 of $\sim100$, with at 
least a factor of 
$\sim6$ uncertainty accounting for our uncertainty of the dust temperature 
($\Delta\,T_\mathrm{d} \simeq \pm{5}$\,K), dust emissivity coefficient 
($\Delta\,\beta \simeq\pm{0.5}$), and mass absorption coefficient 
(about a factor of $\sim3$; e.g.~\citealt{Seaquist04}).

In terms of the spatial distribution of the star formation, if the
24\,$\mu$m emission traces the $R$-band light, then we expect that the
UV-bright component A$_{\rm s}$ will contain 75 per cent of the current
star formation.  Given the poor spatial resolution of our 24\,$\mu$m
image and its modest S/N it is impossible to determine if the emission
follows the UV or CO morphology of the source.  However, we have
confirmed that within the large uncertainty, the integrated
$R/24\,\mu$m color of the LBG is similar to that seen in other $z=3$
LBGs \citep{Reddy06}, supporting the assumption that the 24\,$\mu$m 
follows the $R$-band light.

Assuming the molecular gas reservoir we detect is fueling the
star-formation within this galaxy, then it has enough gas to sustain the
current star-formation for $\tau_{\rm depletion}\sim
$M$(\mathrm{H}_{2})/\mathrm{SFR}\sim 2.4 \times
10^{9}$\,M$_{\odot}/60$\,M$_{\odot}\mathrm{yr}^{-1}\sim40$\,Myrs.
Since we also have a stellar mass of LBG\,J213512, we can compare the
gas depletion time with the time to form the current stellar mass of
the system.  At the current SFR: $\tau_{\rm formation}\sim $M$_{\rm
stars}/\mathrm{SFR}\sim 6 \times
10^{9}$\,M$_{\odot}/60$\,M$_{\odot}\,\mathrm{yr}^{-1}\sim100$\,Myr,
which is comparable to the assumed age of the stellar population.

\section{Discussion}\label{sec:discussion}

Our observations of LBG\,J213512 demonstrate that it is a
relatively massive galaxy hosting an equally
massive gas reservoir and has significant on-going star formation.
The majority of the baryons in the central regions of the galaxy are 
in the form of stars.  We estimate that the stellar population
of the LBG could have been formed at the current SFR within $\sim
100$\,Myrs and that if the current SFR continues, then the cold gas
reservoir will be exhausted within $\lsim 40$\,Myrs unless it is
replenished.  Thus, it appears that we are seeing the LBG in the 
last half of its current star formation episode.  We now compare
the gas properties of LBG\,J213512 to other similarly well-studied
sources and populations at low and high redshifts.

\subsection{Comparison to other populations}

We first compare LBG\,J213512 to the only other CO-detected LBG: cB58
\citet{Baker04}.  The intrinsic CO line luminosity, gas and dynamical
masses we derive for LBG\,J213512 are
L$'_\mathrm{CO(3-2)}=3.0\times10^{9}\,\mathrm{K\,km\,s^{-1}\,pc^{2}}$,
M$_\mathrm{gas}=2.5\times10^{9}$\,M$_{\odot}$ and
M$_\mathrm{dyn}(<1$\,kpc$)=8.4\times10^{9} {\rm csc}^2 i$\,M$_{\odot}$,
with a gas fraction within 1-kpc of $f_{\rm gas}=0.30 \sin^2i$.  Using
the same conversion factors for cB58, $\alpha=0.8$ and $r_{32}=1$, we
estimate:
L$'_\mathrm{CO(3-2)}=0.43\,\times10^{9}\,\mathrm{K\,km\,s^{-1}\,pc^{2}}$,
M$_\mathrm{gas}=0.34\times10^{9}$\,M$_{\odot}$ and
M$_\mathrm{dyn}\sim 10\times10^{9} {\rm csc}^2 i$\,M$_{\odot}$ (which
has FWHM$_{\rm CO}= 175\pm 45$\,km\,s$^{-1}$), with a gas fraction of
$f_{\rm gas}= 0.03\sin^2i$.

We could attempt to reduce the uncertainties due to the unknown
inclination and average the properties of LBG\,J213512 and cB58,
deriving a typical dynamical mass of M$_\mathrm{dyn}\sim
(3.6\pm0.4)\times10^{10}$\,M$_{\odot}$ and a gas fraction of $f_{\rm
gas}\sim 0.05\pm 0.05$ (both for $i=30$\,degrees).  However, we note
that LBG\,J213512 has roughly $7\times$ more cold gas than cB58,
demonstrating that there is a significant variation in the gas content
of $L^\ast$ LBGs at $z\sim 3$ and argues against blindly averaging their
properties.  Our estimate of the SFR for LBG\,J213512 is 3--$4\times$
higher than cB58's value of 24\,M$_\odot$\,yr$^{-1}$ \citep{Baker04}, 
although this comparison uses different indicators in the two
LBGs and so is uncertain.  These SFRs suggest comparable gas depletion
time scales for LBG\,J213512 and cB58: $\tau_{\rm depletion} \lsim
40$\,Myrs and $\tau_{\rm
depletion}\sim0.34\times10^{9}/24\sim15$\,Myrs for cB58.  The
variations in gas mass as well as the short depletion timescales
probably reflects the brevity of the star formation events and hence
the strong variation within the population in the amount of gas
consumed at any time.  In terms of the comparison of LBG\,J213512 and
cB58: these appear to be similarly massive L$^\ast$ LBGs, although
LBG\,J213512 is younger, more gas-rich and forming stars at a higher
rate than cB58.

The only high-redshift galaxy population for which there are large
numbers of sources with reliable cold gas masses are submillimeter
galaxies (SMGs).  Taking the average CO line luminosity and gas mass
for cB58 and LBG\,J213512 we find that the ``typical'' LBG has an
intrinsic line luminosity $\sim 20\times$ lower than the median for SMGs
(c.f.\ $<\!L'_\mathrm{CO}\!>=(3.8\,\pm{2.3})\times10^{10}$ and
$<\!M_\mathrm{gas}\!>=(3.0\,\pm{1.6})\times10^{10}$;
\citealt{Greve05}).  Although, we note that the SMGs studied by
\citet{Greve05} are not strongly lensed and hence the sensitivity
limits of their survey precludes the detection of sources with
L$'_\mathrm{CO}\lsim 1\times10^{10}\,\mathrm{K\,km\,s^{-1}\,pc^{2}}$,
and that in addition, the typical SMGs lies at somewhat lower redshifts
than the LBGs (although those SMGs at $z\sim 3$ appear comparable to
the general population).  The CO line widths of LBG\,J213512 and cB58
are $3\times$ lower than the median of SMGs (\citealt{Greve05}) with
this emission arising in a region estimated to be half the size of that
in SMGs (\citealt{Tacconi06}).  The typical gas-to-dynamical mass
fraction in SMGs is estimated to be $\sim0.3$ assuming a merger model
\citep{Greve05}, while they have SFEs of $L_{\rm
  FIR}/M_\mathrm{H_2}\sim 450\pm 170$\,L$_\odot$\,M$_\odot^{-1}$
\citep{Greve05}, gas-to-dust mass ratios of $\sim200$ 
(with about factor of a few uncertainty in the dust mass alone)
and gas surface densities of
$\Sigma_{gas}\sim3000$\,M$_{\odot}$\,yr$^{-1}$\,pc$^{2}$
\citep{Tacconi06}.  

Overall, this comparison suggests that LBGs and SMGs are
equally evolved, but that the cold gas reservoir in LBGs resides in a
system which is typically a factor of $\sim 5\times$ less massive than
in SMGs \citep{Swinbank04,Swinbank06}, that the LBG's cold gas disks
have surface densities $4\times$ lower than SMGs (if both can be
well-described as disks) and that LBGs appear to be forming stars less
efficiently than typical SMGs, by about a factor of $4\times$.  
The two populations do appear to have similar
fractions of baryons in cold gas and stars: M$_{\rm gas}/$M$_{\rm
  stars} \sim 0.2$, 0.4 and $\sim 0.3$ for cB58, LBG\,J213512 and SMGs,
respectively. 

We can also compare the LBGs to local populations.  We first note that
our 24-$\mu$m detection predicts a far-infrared luminosity for
LBG\,J213512 of $L_{\rm FIR}\sim 3\times 10^{11}L_\odot$.  Even with
our estimated factor of $3\times$ uncertainty, this indicates that the
LBG is likely to have a far-infrared luminosity comparable to local
LIRGs, $L_{\rm FIR}\gsim 10^{11}L_\odot$.  

Locally, L$'_{\rm CO}$ increases with L$_{\rm FIR}$ for (U)LIRGs, with the 
\citet{Greve05} sample of SMGs extending this 
trend out to the highest far-infrared luminosities 
($\gtrsim10^{13}$L$_{\odot}$).  
For comparison, in Fig.~3 we have plotted LBG\,J213512 on the 
L$'_{\rm CO}$--L$_{\rm FIR}$ diagram along with cB58, SMMJ16359+6612
(with a lensing magnification of $22\times$ and weak submillimeter emission, 
making it more similar to LBG\,J213512 than to the SMG population) from 
\citet{Kneib05}, LIRGs, ULIRGs and SMGs.  LBG\,J213512, cB58 and 
SMMJ16359+6612 all lie on the local relation within the considerable 
uncertainties in their far-infrared luminosities, in the same region 
of the plot that the local LIRGs occupy.  This suggests that LBG\,J213512 
is similar to local LIRGs.

The CO line widths (and hence dynamical masses for similar radii 
and inclination angles)
for LIRGs ($<\!{\mathrm{FWHM}}\!>\sim200\,\mathrm{km\,s^{-1}}$;
\citealt{Sanders91}) are comparable to LBG\,J213512 and cB58,
suggesting that LBGs are simply gas-rich high-redshift analogs of
LIRGs, but possessing a marginally higher star formation efficiency
than the range typically seen in local LIRGs 
(140\,L$_\odot$\,M$_\odot^{-1}$ compared to $\sim
1$--50\,L$_\odot$\,M$_\odot^{-1}$ for $\alpha=0.8$; \citealt{Sanders91}).

\begin{inlinefigure}
  \centerline{\psfig{file=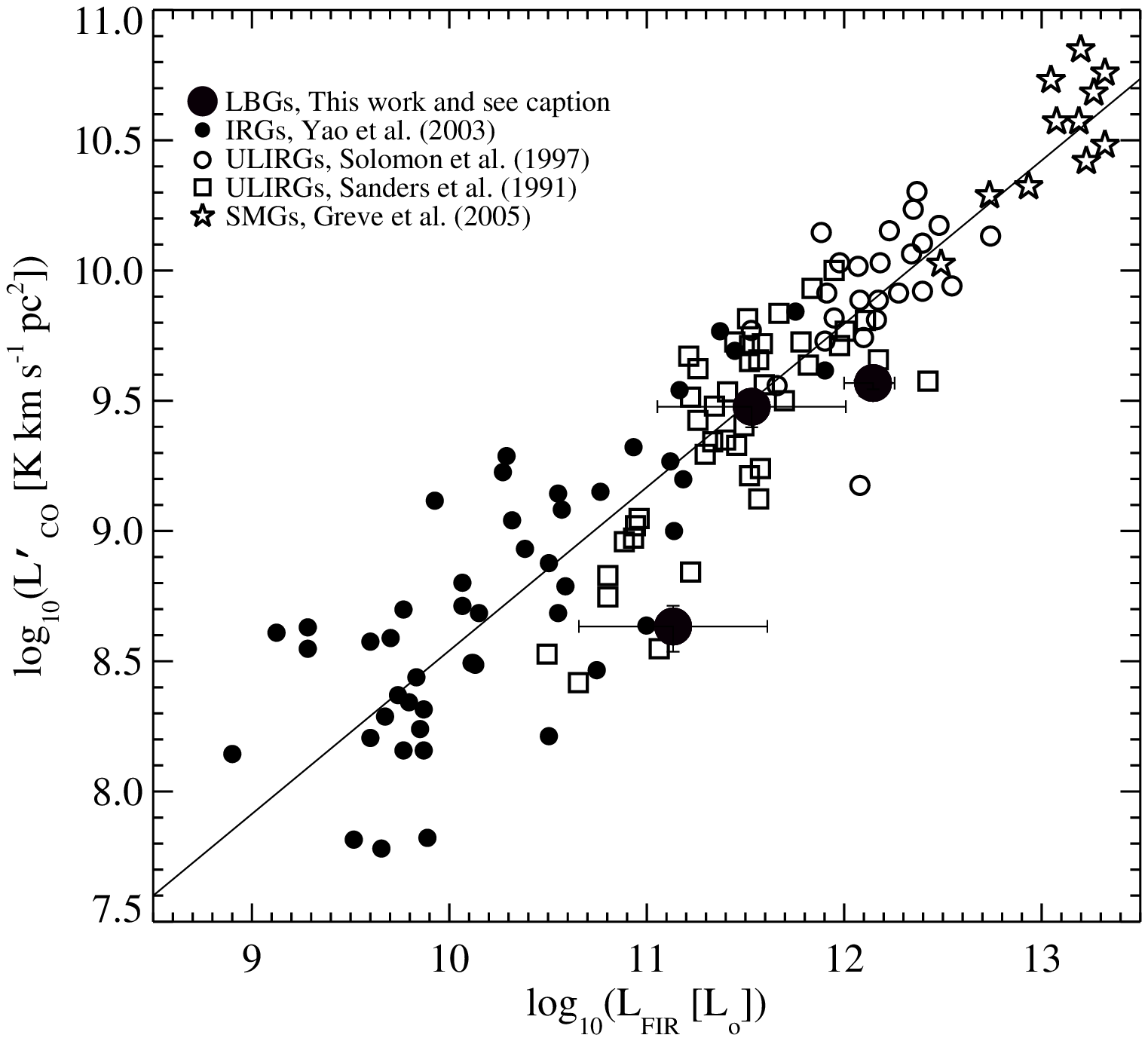,width=3.5in,angle=0}}
  \figcaption{A comparison of the CO and far-infrared luminosities for LBGs 
(including cB58 from \citealt{Baker04} and LBG\,J213512), and the faint 
submillimeter-selected source SMMJ16359+6612 from \citealt{Kneib05} 
(listed in order of increasing luminosity), 
LIRGs, ULIRGs and SMGs.  The solid line is the best-fitting relation 
with a form of 
log L$'_{\rm CO}=\alpha\,\mathrm{log}$\,L$_\mathrm{FIR}+\beta$ 
to the LIRGs, ULIRGs and SMGs from \citet{Greve05}.
The LBGs all lie on the relation within their 
uncertainties and lie in the same region of the plot as the LIRGs.  
Note that this diagram is a particularly useful diagnostic since it does not 
depend on the CO-to-H$_{2}$ conversion factor.}
\label{fig:lcolfir}
\end{inlinefigure}
\addtocounter{figure}{0}

\subsection{The physics of star-formation in LBGs}

What do our observations tell us about the structure of the gas
disk and the mode of star formation in LBGs?

To investigate the physics of star formation within
LBGs in more detail, 
we can first test whether the
star formation activity within LBG\,J213512 is consistent with the global
Schmidt law \citep{Schmidt59}.
\citet{Kennicutt98} derives a Schmidt law of the form 
$\Sigma_\mathrm{SFR}=A\,(\Sigma_\mathrm{gas})^{1.4}$,
where $\Sigma_\mathrm{gas}$ and $\Sigma_\mathrm{SFR}$ are in units of
$M_\odot\mathrm{yr}^{-1}\mathrm{pc}^{-2}$ and
$M_\odot\mathrm{yr}^{-1}\mathrm{kpc}^{-2}$, respectively and
$A=(2.5\pm0.7)\times10^{-4}$.  
In deriving this relationship,
\citet{Kennicutt98}
assumed  $\alpha=4.6$  for {\it all} the galaxies (including
LIRGs) and noted that the scatter around the relation is large, with
individual galaxies lying up to a factor of $7\times$ off the mean relation. 
  
Therefore, we follow \citet{Kennicutt98} and use $\alpha=4.6$ 
for  the gas surface densities in this calculation.
For LBG\,J213512 we 
estimate a normalisation of $A=1.6^{+3.2}_{-1.1}\times10^{-4}$,
where the large error accounts for the  uncertainty in
the true SFR.  In contrast, cB58 predicts a normalization of
$A\sim9.3\times10^{-4}$ with a similar uncertainty.  
Thus the two LBGs bracket the local Schmidt law and are
each consistent with it within the large scatter and uncertainties. 

To determine the likely structure of the gas reservoir, we use the gas
mass and assume a radius of 1\,kpc to infer a mass surface density of
$\Sigma=760\pm130\,\mathrm{M_\odot}$\,pc$^{-2}$.  
If the gas is present in a disk, we
can test whether the material will be unstable to bar-formation.  The
Toomre stability criterion states that $Q={\sigma\kappa}/{2\pi
  G\Sigma}\gsim1$ in order for a gaseous disk to be stable
(e.g.~\citealt{BinneyTremaine}) where $\sigma$ and $\kappa$ are the
velocity dispersion and the circular frequency of the disk
respectively.  For a surface density of 760\,M$_\odot$\,pc$^{-2}$, 
$Q<1$ for any realistic value of $\sigma$ and $\kappa$.  
This suggests that if the gas is
present in a disk, it will be unstable to bar-formation and will
collapse on a timescale of  $\tau_{\mathrm{rot}}\sim60$\,Myrs,
comparable to the estimated remaining lifetime of the
burst.

Following \citet{Tacconi06}, we derive the maximal SFR and SFR surface
density for the average LBG with an inclination angle of
$i=30\,\mathrm{deg}$ to be $\sim$20\,M$_\odot\,\mathrm{yr}^{-1}$ and
$\sim$7\,M$_\odot\,\mathrm{yr}^{-1}\,\mathrm{kpc}^{-2}$, respectively.
These are comparable to the observed average SFR and SFR surface
density of $\sim$40\,M$_\odot\,\mathrm{yr}^{-1}$ and
$\sim$7\,M$_\odot\,\mathrm{yr}^{-1}\,\mathrm{kpc}^{-2}$, respectively. 
Hence we conclude that LBGs could be maximal starbursts if our
assumptions of the geometry and star-formation efficiency are all
correct.  Our observations appear to point towards the star formation in
LBG\,J213512 occuring in an intense central starburst, similar to
the activity seen in local LIRGs and ULIRGs.  Many of
the physical properties we derive for the LBG mirror those
of local LIRGs.  However, many of 
these conclusions are sensitive to the conversion factor between
the CO and H$_2$ gas masses, $\alpha$, and so we must
re-examine this choice.

The values of
$\alpha=0.8$\,M$_{\odot}(\mathrm{K\,km\,s^{-1}\,pc^{2}})^{-1}$ 
and $r_{32}=1$
we adopted are derived from luminous local starburst galaxies  
in which the molecular gas is distributed in a extensive
intercloud medium  \citep{S097}, rather than as discrete
giant molecular clouds (GMCs).  Thus lower
H$_2$ to CO ratios in these more energetic environments are potentially
an indication of a mode of star formation associated with bulge formation,
with higher ratios being more prevalent in quiescently star-forming disks.
A value of $\alpha=0.8$
is also adopted for high-redshift submillimeter 
galaxies (which are believed to be scaled-up high-redshift analogs of
local ULIRGs; \citealt{Greve05,Tacconi06}). If we had instead assumed
that the cold gas is distributed in stable GMCs, as in our own galaxy,
then the predicted gas masses are  $9\times$ higher
(taking $\alpha=4.6$\,M$_{\odot}(\mathrm{K\,km\,s^{-1}\,pc^{2}})^{-1}$ 
and $r_{32}=0.65$; \citealt{SolomonBarrett}; \citealt{Mauersberger99}; 
\citealt{Dumke01}; \citealt{Yao03}).  

We currently cannot constrain $\alpha$ in high redshift galaxies directly
and so we must examine whether
adopting the Milky Way conversion factors would substantially alter our
interpretation of the LBGs (c.f.~\citealt{Baker04}).  Adopting
$\alpha=4.6$ and $r_{32}=0.65$ increases the gas masses for
LBG\,J213512 to $22\times 10^{9}$\,M$_\odot$, which is comparable to
the cold gas masses derived for typical SMGs \citep{Greve05} and
exceeds the dynamical mass of the galaxy for inclinations 
of $i\gsim 40$\,deg.  Similarly cB58's gas mass increases 
to $3\times 10^{9}$\,M$_\odot$.  The star
formation efficiency, L$_{\rm FIR}/$M$_{\rm H_2}$, for LBG\,J213512
drops to $\sim 15$L$_\odot$\,M$_\odot^{-1}$, while the surface density of
the gas disk increases dramatically to
$\Sigma_{\mathrm{gas}}\sim8000$\,M$_\odot$\,pc$^{-2}$, larger than the
values claimed for SMGs \citep{Tacconi06}.  

It is difficult to understand this mix of
properties, with immense gas reservoirs in highly unstable, dense gas
disks which are in some respects more extreme than SMGs, but yet result
in star formation occuring with an efficiency at least an order of
magnitude below that in SMGs.  We therefore conclude that it is
hard to reconcile the gas properties of LBGs if the CO-to-H$_2$
conversion factor is similar to the Milky Way.  Instead, we suggest
that the high gas surface density and low star-formation efficiency is
most naturally explained if the gas is distributed in a similar manner
to local ULIRGs and SMGs and the dominant star-formation mode 
follows that seen in local luminous starbursts (e.g.\ ULIRGs).

However, the cost of adopting $\alpha=0.8$ is that the gas fraction
for cB58 drops to just $f_{\rm gas}\sim 0.03\sin^2i$, which seems
uncomfortably low for a young high-redshift galaxy which is still
forming stars.  One potential solution is to adopt different 
CO-to-H$_2$ conversion factors in the two LBGs, with the more
active LBG\,J213512 having a
lower value than cB58.  This could reflect a period of bulge-formation
in LBG\,J213512, while cB58 is currently in a less active
phase of star formation.   Such a complication is of course unjustified
given the current information we have and we must instead look forward
to observations of cold gas in a larger number of LBGs to allow us to
place limits on $\alpha$ using the same dynamical arguments 
employed by \citet{S097}.

\section{Conclusions}

We have carried out millimeter interferometry and mid-infrared imaging
of a strongly lensed LBG at $z=3.07$.  We detect strong CO(3--2)
emission with a line width of $190\pm24$\,km\,s$^{-1}$.
Although the lensing magnification of the entire UV source is $\sim
28 \times$, the position of the CO emission appears coincident with a
UV component in the source plane which has a magnification of
$\sim8\times$.  Correcting for the magnification of the
CO source, LBG\,J213512 has an inferred gas mass of $(2.4 \pm
0.4)\times10^{9}$\,M$_{\odot}$ and a dynamical mass of $(8.4 \pm1.4)
\times10^{9}\,{\rm csc}^{2}i$\,M$_\odot$, within an estimated radius of
$\sim 1$\,kpc.
Fitting to the observed $gR_{606}K$ and our \textit{Spitzer} IRAC
3.6--8.0\,$\mu$m photometry we derive a stellar mass of $\sim (6\pm 2)
\times 10^{9}$\,M$_{\odot}$.  
We also use {\it Spitzer}/MIPS
$24\mu$m imaging to estimate a 
far-infrared luminosity of $\sim 3\times10^{11}$\,L$_\odot$ with an
uncertainty of a factor of $3\times$.  From this we derive a 
star-formation rate of
$\sim60$\,M$_\odot$\,yr$^{-1}$, and hence a
star-formation efficiency of $\sim 140$\,L$_\odot $\,M$_\odot^{-1}$. 
Based on this star formation rate we estimate that the current activity
in LBG\,J213512 could have formed the current stellar mass
in a period of $\sim 100$\,Myrs and 
can continue for a further $\sim 40$\,Myrs before the gas
reservoir is exhausted.  

We find that the CO line luminosity and inferred gas mass for
LBG\,J213512 are $\sim 7\times$ higher than that
measured for the only other CO-detected LBG, cB58, indicating a
significant variation in the gas content of L$^\star$ LBGs at
$z\sim3$.  In contrast, the two LBGs have effectively
identical CO line-widths, indicating similar dynamical masses (within
the uncertainties of their unknown inclinations).
Thus there is a large range in gas-to-dynamical mass ratio between
cB58 and LBG\,J213512,
$M_{\mathrm{gas}}/M_{\mathrm{dyn}}\sim$\,0.03--$0.3\sin^2 i$, which may 
reflect the short timescales for star formation in these galaxies.
Following Kennicutt (1998), we use the gas and star-formation
surface densities to derive the Schmidt law and find that both
LBG\,J213512 and cB58 are consistent with the Schmidt law.  

We compare our observations of LBG\,J213512 to far-infrared
luminous galaxy populations at low and high redshifts.  We
find that 
the star-formation efficiency we derive for LBG\,J213512 is
slightly above the range of values derived for similarly
far-infrared-luminous local galaxies (LIRGs), but is
significantly below the typical efficiency for
high-redshift submillimeter galaxies
($\sim450$\,L$_\odot$\,M$_\odot^{-1}$).  The gas and dynamical
masses we find are also an order of magnitude
smaller than estimated for SMGs.  We suggest that LBG\,J213512
has many features in common with local LIRGs and hence
that the star formation
activity in this high-redshift system may be closely related
to the mode of star formation seen in these galaxies.

A major uncertainty in our analysis is the choice of the 
CO-to-H$_2$ conversion factor ($\alpha$) we should adopt.  The two most
commonly used assumptions: $\alpha=0.8$ and $r_{32}=1$ appropriate for local
ULIRGs or $\alpha=4.6$ and $r_{32}=0.65$ as measured for the 
Milky Way, produce a factor of $\sim 9\times$ variation 
in the expected gas masses
and all the associated derived properties.  We argue that
selecting an $\alpha$ and $r_{32}$ similar to comparably far-infrared luminous
galaxies at the present-day yields a system with more easily
understood properties, than the more gas-rich system predicted
by $\alpha=4.6$ and $r_{32}=1$.  However, 
it is clear that a measurement of other CO transitions for LBG\,J213512, 
accessible from the GBT or from the PdB, 
would place a constraint on the temperature and density of the
molecular gas (e.g.~\citealt{Hainline_oct06}) and allow a more
accurate determination of the line luminosity ratio, $r_{32}$, and
hence the total gas mass of the system.  Placing limits on the
CO-to-H$_2$ conversion factor, $\alpha$, in LBGs will require either
high
angular resolution CO mapping (to resolve the gas disks and solve for
their inclination), which could be within the grasp of the
most extended IRAM configurations, or from a statistical study
of the gas-to-dynamical mass ratios for a large sample of LBGs with
CO detections.  The latter may have to wait until the advent of ALMA.

Overall, these observations effectively harness a gravitational lens to
boost the light-grasp of the IRAM and \textit{Spitzer} telescopes by a
factor of up to $30\times$, whilst improving the effective resolution to
$\sim 0.2\arcsec$.  The next step is to observe the system at 
higher resolution with 
IRAM (FWHM $\sim0.3\arcsec$) in order to dissect
the gas distribution on the smallest scales of $\sim0.02\arcsec$ 
(corresponding to $\sim0.2\,\mathrm{kpc}$ in the image plane),
providing a preview of the capabilities of ALMA.  Such observations
will allow us to address questions which are drivers of the ALMA, SKA
and {\it JWST} science cases: yielding insights into the kinematics of
the interstellar medium in a normal, young galaxy seen 12 billion years
ago.

\acknowledgements
This work is based on observations carried out with the IRAM PdBI.
IRAM is supported by INSU/CNRS (France), MPG (Germany) and IGN (Spain).
This work is also makes use of data obtained with \textit{Spitzer},
which is operated by the Jet Propulsion Laboratory, California
Institute of Technology under a contract with NASA.  We thank the
\textit{Spitzer} Director, Tom Soifer, for the award of the
\textit{Spitzer} DDT.  We thank Max Pettini, Alice Shapley, 
and Dave Alexander for useful discussions and help, 
and an anonymous referee for a constructive report.
KEKC, AMS and JEG acknowledge support from PPARC.  IRS acknowledges
support from the Royal Society.

\clearpage


\begin{thebibliography}{}
\bibitem[Baker et al.(2004)]{Baker04}Baker A.J., Tacconi L.J., Genzel R., Lehnert D., Lutz D., 2004, \apj, 604, 125
\bibitem[Baugh et al.(2005)]{Baugh05}Baugh C.M., Lacey C.G., Frenk C.S., Granato G.L., Silva L., Bressan, Benson A.J., Cole S., 2005, \mnras, 356, 1191
\bibitem[Bell et al.(2005)]{Bell05}Bell E.F. et al., 2005, ApJ, 625, 23
\bibitem[Binney \& Tremaine(1987)]{BinneyTremaine}Binney J., Tremaine S., 1987, Galactic Dynamics, Princeton Univ. Press, Princeton, NJ
\bibitem[Blain et al.(2002)]{Blain02}Blain A.W., Smail I., Ivison R.J., Kneib J.-P., Frayer D.T., 2002, Phys. Rep., 369, 111
\bibitem[Bolzonella, Miralles \& Pello(2000)]{hyperz} Bolzonella M., Miralles J.-M., Pello R., 2000, A\&A, 363, 476
\bibitem[Borys et al.(2005)]{Borys05} Borys C., Smail I., Chapman S.C., Blain A.W., Alexander D.M., Ivison R.J., 2005, \apj, 635, 853
\bibitem[Bruzual \& Charlot(1993)]{BruzChar93}Bruzual G., Charlot S., 1993, \apj, 405, 538
\bibitem[Calzetti et al.(2000)]{Calzetti00}Calzetti D., Armus L., Bohlin R.C., Kinney A.L., Koornneef J., Storchi-Bergmann T., 2000, \apj, 533, 682
\bibitem[Carilli \& Wang(2006)]{CarWangErr}Carilli C.L., Wang R., 2006, \aj, 132, 2231
\bibitem[Chapman et al.(2000)]{Chapman00}Chapman S.C. et al., 2000, \mnras, 319, 318
\bibitem[Dale \& Helou(2002)]{DaleHelou02}Dale D.A., Helou G., 2002, \apj, 576, 159
\bibitem[Dale et al.(2001)]{Dale01}Dale D.A., Helou G., Contursi A., Silbermann N.A.,  Kolhatkar S., 2001, \apj, 549, 215
\bibitem[Dumke et al.(2001)]{Dumke01}Dumke K., Nieten C., Thuma R., Wielebinski R., Walsh W., 2001, A\&A, 373, 853
\bibitem[Dye et al.(2007)]{Dye}Dye S., Smail I., Swinbank A.M., Warren S.J., Ebeling H., Edge A.C., 2007, \apj (submitted)
\bibitem[Geach et al.(2006)]{Geach06} Geach J.E. et al., 2006, \apj, 649, 661
\bibitem[Greve et al.(2005)]{Greve05} Greve T.R. et al., 2005, \mnras, 359, 1165
\bibitem[Guilloteau et al.(1992)]{Guilloteau}Guilloteau S. et al., 1992, A\&A, 262, 624
\bibitem[Hainline et al.(2006)]{Hainline_oct06} Hainline L.J., Blain A.W., Greve T.R., Chapman S.C., Smail I., Ivison R.J., 2006, \apj, 650, 614
\bibitem[Kennicutt(1998)]{Kennicutt98}Kennicutt R.C. Jr., 1998, \apj, 498, 541
\bibitem[Kneib et al.(2005)]{Kneib05}Kneib J.-P., Neri R., Smail I., Blain A., Sheth K., van der Werf P., Knudsen K.K., 2005, A\&A, 434, 819
\bibitem[Law et al.(2007)]{Law07}Law D.L., Steidel C.C., Erb D.K., Pettini M., Reddy N.A., Shapley A.E., Adelberger K.L., Simenc D.J., 2007, ApJ, 656, 1
\bibitem[Leitherer(1999)]{Leitherer99}Leitherer C., et al., 1999, \apjs, 123, 3L
\bibitem[Mauersberger et al.(1999)]{Mauersberger99}Mauersberger R., Henkel C., Walsh W., Schulz A., 1999, A\&A, 341, 256
\bibitem[Pettini et al.(2000)]{Pettini00}Pettini M., Steidel C.C., Adelberger K.L., Dickinson M., Giavalisco M., 2000, \apj, 528, 96
\bibitem[Pettini et al.(2002)]{Pettini02}Pettini M., Rix S.A., Steidel C.C., Adelberger K.L., Hunt M.P., Shapley A.E., , 2002, \apj, 569, 742
\bibitem[Reddy \& Steidel(2004)]{Reddy04}Reddy N.A., Steidel C.C., 2004, \apj, 603, L13
\bibitem[Reddy et al.(2006)]{Reddy06}Reddy N.A., Steidel C.C., 2006, Fadda D., Yan L., Pettini M., Shapley A.E., Erb D.K., Adelberger K.L., \apj, 644, 792
\bibitem[Sanders, Scoville \& Soifer(1991)]{Sanders91}Sanders D.B., Scoville N.Z., Soifer B.T., 1991, \apj, 370, 158
\bibitem[Schmidt(1959)]{Schmidt59}Schmidt M., 1959, \apj, 129, 243
\bibitem[Seaquist et al.(2004)]{Seaquist04} Seaquist E., Yao L., Dunne L., Cameron H., 2004, \mnras, 349, 1428
\bibitem[Shapley et al.(2001)]{Shapley01}Shapley A.E., Steidel C.C., Adelberger K.L., Dickinson M., Giavalisco M., Pettini M., 2001, \apj, 562, 95
\bibitem[Shapley et al.(2003)]{Shapley03}Shapley A.E., Steidel C.C., Pettini M., Adelberger K.L., 2003, \apj, 588, 65
\bibitem[Shapley et al.(2005)]{Shapley05}Shapley A.E., Steidel C.C., Erb D., Reddy N.A., Adelberger K.L., Pettini M., Barmby P., Huang J., 2005, \apj, 626, 698
\bibitem[Smail et al.(2007)]{Smail06} Smail I. et al., 2007, \apj, 654, 33
\bibitem[Solomon \& Barrett(1991)]{SolomonBarrett} Solomon P.M., Barrett J.W., 1991, In \textit{Dynamics of Galaxies and Their Molecular Cloud Distributions:  Proceedings of the 146th Symposium of the IAU}, 146th, ed. F. Combes, F.Casoli, pp. 235-241.  Dordrecht:  Kluwer
\bibitem[Solomon \& Vanden Bout(2005)]{Solomon_rev} Solomon P.M., Vanden Bout P.A., 2005, \araa, 43, 677
\bibitem[Solomon et al.(1997)]{S097} Solomon P.M., Downes D., Radford S.J.E., Barrett J.W., 1997, \apj, 478, 144
\bibitem[Somerville, Primack \& Faber(2001)]{Somerville01} Somerville R.S., Primack J.R., Faber S.M., 2001, \mnras, 320, 504
\bibitem[Spergel et al.(2003)]{Spergel}Spergel D.N. et al., 2003, \apjs, 148, 175
\bibitem[Steidel et al.(1996)]{Steidel96}Steidel C.C., Giavalisco M., Pettini M., Dickinson M., Adelberger K.L., 1996, \apj, 462, L17
\bibitem[Swinbank et al. (2004)]{Swinbank04}Swinbank, A. M., Smail, I., Chapman, S. C., Blain, A. W., Ivison, R. J., Keel, W. C., \apj, 617, 64
\bibitem[Swinbank et al. (2006)]{Swinbank06} Swinbank, A. M., Chapman, S. C., Smail, I., Lindner, C., Borys, C., Blain, A. W., Ivison R. J., Lewis, G. F., 2006 \mnras 371, 465
\bibitem[Tacconi et al.(2006)]{Tacconi06}Tacconi L.J. et al., 2006, \apj, 640, 228
\bibitem[Wall \& Jenkins(2004)]{Wall}Wall J.V., Jenkins C.R., 2003, Practical Statistics for Astronomers.  Cambridge Univ. Press, Cambridge
\bibitem[Yao et al.(2003)]{Yao03}Yao L., Seaquist E.R., Kuno N., Dunne L., 2003, \apj, 588, 771
\bibitem[Yee et al.(1996)]{Yee96}Yee H.K.C., Ellingson E., Carlberg R.G., 1996, \apjs, 102, 269
\end{thebibliography}
\end{document}